# Ice Formation in Model Biological Membranes in the Presence of Cryoprotectors

M. A. Kiselev*, P. Lesieur[#], A. M. Kisselev[$] and M. Ollivon[+]

* L.U.R.E., Université Paris-Sud, Bât. 209-D, F91405 Orsay cedex, France.
  Permanent address: Frank Laboratory of Neutron Physics, JINR, Dubna 141980, Moscow region, Russia
[#] C.E.A., Centre de Saclay, DRECAM-SCM, F-91191 Gif-sur-Yvette cedex, France
[$] Moscow Insitute of Physics and Technology, Dolgoprudniy 141700, Russia
[+] Physico-Chimie des systèmes polyphasés, URA 1218 du CNRS, Faculté de Pharmacie, tour B, F – 92296, Chatenay Malabry, France

**ABSTRACT**

Ice formation in model biological membranes is studied by SAXS and WAXS in the presence of cryoprotectors: dimethyl sulfoxide and glycerol. Three types of phospholipid membranes: DPPC, DMPC, DSPC are chosen for the investigation as well-studied model biological membranes. A special cryostat is used for sample cooling from 14.1°C to −55.4°C. The ice formation is only detected by WAXS in binary phospholipid/water and ternary phospholipid/cryoprotector/water systems in the condition of excess solvent. Ice formation in a binary phospholipid/water system creates an abrupt decrease of the membrane repeat distance by $\Delta d$, so-called ice-induced dehydration of intermembrane space. The value of $\Delta d$ decreases as the cryoprotector concentration increases. The formation of ice does not influence the membrane structure ($\Delta d = 0$) for cryoprotector mole fractions higher than 0.05.

PACS codes: 87.15.By, 87.15.Kg, 87.22.Bt, 61.10.Eq

Keywords: Dimethyl sulfoxide, glycerol, cryoprotection, phospholipid, X-ray scattering.

*Correspondence to* :
M.A. Kiselev - Frank Laboratory of Neutron Physics, Joint Institute for Nuclear Research, 141980 Dubna, Moscow region, Russia.
E-mail: kiselev@nf.jinr.ru;  fax: 7-096-21-65882; phone: 7-096-21-66275.

## 1. Introduction

Dimethyl sulfoxide (DMSO) and glycerol are well-known cryoprotectors widely used in cryobiology for the conservation of biological tissues at low temperatures [1]. The main purpose of cryobiology is to find an "optimal" way for cooling biological systems to low temperatures (about the liquid nitrogen temperature), and prevent the crystallisation of water inside biological tissues. One of the problems is a longer time of freeze-induced dehydration (diffusion of water molecules from biological tissues through the membrane to the conserving solvent) than the time of ice formation [2]. The ice-induced dehydration of intermembrane space at the moment of water crystallization is observed by X-ray diffraction [3] and calorimetry [4] for binary phospholipid /water systems. In macromolecular crystallography the portion of X-ray diffraction experiments at cryogenic temperatures is exponentially increasing. In many cases, a dramatic reduction in radiation damage at low temperatures allows complete data sets to be collected from a single crystal. cryoprotectors make it possible to cool samples to low temperature and store them without damage in the crystal [5].

In this article ice formation is investigated by small- and wide-angle X-ray scattering (SAXS and WAXS) in ternary phospholipid/cryoprotector/water systems. Three phospholipids are studied: dimyristoylphosphatidylcholine (DMPC), dipalmitoylphosphatidylcholine (DPPC) and distearoylphosphatidylcholine (DSPC). DPPC, DMPC, and DSPC are taken as phospholipids with an equal polar heads and with the difference in the length of hydrocarbon chains. The reason why we use DMSO and glycerol is the fact of their different interactions with the membrane surface. Glycerol penetrates into the region of polar head groups and produces influence on the thickness of the membrane bilayer and the thickness of an intermembrane solvent [6]. DMSO molecules do not penetrate into the region of polar head groups and thus do not influence the membrane thickness [7,8]. DMSO molecules strongly interact with water molecules and form hydrogen bonds. Such interaction is more intensive than the interaction between DMSO and the polar head groups of DPPC [9]. Figure 1 presents a model water topology in the intermembrane space of a `phospholipid/water system [4, 10].

## 2. EXPERIMENT

DPPC, DMPC, DSPC (over 99% pure), spectrophotometric grade DMSO and glycerol are purchased from Sigma and Aldrich corporations (Paris, France). Water (18 MΩ·cm) are obtained by means of Millipore. The samples are prepared in quartz capillaries. The capillaries (1.5 mm in diameter and 0.01 mm wall thickness) are purchased from GLAS company (W.Muller, Berlin, Germany).

X-ray diffraction patterns are measured at LURE on the D22 beam line of DCI synchrotron ring. The time resolved diffraction is used to collect information about ice formation in the studied systems. Two linear position sensitive detectors are used. The first detector is placed at sample to detector distance of 1690 mm in order to record the diffraction from a lipid multilamellar structure in the region of the scattering vectors $-0.3 \text{Å}^{-1} \leq q \leq 0.3 \text{ Å}^{-1}$ (SAXS measurements). The second detector (the sample to detector distance is 362 mm , at angle of $70^o$ to the beam direction) is used to record diffraction patterns from ice and a mutual parking of hydrocarbon chains in the region of the scattering vector $1.3 \text{Å}^{-1} \leq q \leq 1.9 \text{ Å}^{-1}$ (WAXS measurements). Such geometry gives a relatively small uncertainty in the determination of the repeat distance (about 0.5 %). The diffraction patterns are measured at a rate of one minute per frame, the temperature of the sample is changed at a cooling rate of 1.5 degree per minute. The period ***d*** (repeat distance) of multilamellar structures is determined from the position of the first diffraction peak using the Bragg equation: $2 \cdot d \cdot sin\theta = \lambda$, where θ is a half of the scattering angle.

The capillary holder used for X–rays consists of a cryostat developed in Paris–Sud University by G. Keller (CNRS, URA 1218). The sample is cooled by two Peltier elements and heat is evacuated using liquid flow. A resistor is used to regulate the sample temperature and vary the temperature from +14°C to −55.4°C at a constant rate. Great care is taken to prevent water condensation on the surface of the capillary, including both 4 mylar windows and dry nitrogen flow. A thermocouple is used to measure the temperature with an accuracy of ± 0.1°C and a reproducibility of ± 0.5°C.

## 3. RESULTS AND DISCUSSION

Figures 2a and 2b present a sequence of diffraction SAXS and WAXS patterns recorded in the process of sample cooling from $T = +14.1°C$ to $T = -29.5°C$. Figure 2a presents the time resolved diffraction from a multilamellar structure of a DPPC membrane. At cooling from 14.1°C to −16.2°C the diffraction patterns correspond to the transformation of the gel phase in to the crystalline phase [11]. The shift of the diffraction peak in the direction of a larger scattering vector $q$ at $T_{ice} = -19.4°C$ corresponds to a decrease of the membrane repeat distance $d$. It is the result of ice formation in a bulk solvent. Figure 2b demonstrates the formation of three diffraction peaks of ice at $T_{ice} = -19.4°C$ which correspond to the lattice constants: 3.9Å, 3.6Å, 3.4Å. The WAXS from a mutual parking of hydrocarbon chains (see Fig. 2b) at $T = 14.1°C$ shows reflection at $q = 1.49 \pm 0.002$Å$^{-1}$ and a broad shoulder at $q = 1.53 \pm 0.01$ Å$^{-1}$. These reflections correspond to an orthorhombic subcell with the lattice constants 8.46 ± 0.01Å and 9.38± 0.07Å [11]. The two reflections separate monotonously as the temperature decreases to the temperature of ice formation in the positions $1.46 \pm 0.002$Å$^{-1}$ and $1.58 \pm 0.002$Å$^{-1}$ (lattice constants 8.61 ± 0.01Å and 8.97 ± 0.08Å). Slow progressive conversion of interchain space is the result of membrane crystallization. The membrane cystallization is a time and temperature dependent process with a characteristic time of 12 hours [11]. The description of such process is beyond of the scope of this article. Membrane crystallization can be negligible in the study of a fast process, such as water crystallization (character time is 3min). The main goal of our study is to describe changes in the membrane repeat distance $\Delta d$ at the moment of ice formation. The repeat distance of a DPPC multilayer, $d$, is related to the membrane thickness $d_l$ and the thickness of the solvent $d_w$ in the intermembrane space via the expression $d = d_l + d_w$. A decrease in $d$ at $T_{ice} = -19.4°C$ may occur as a result of a decrease in $d_l$ or $d_w$. At the moment of ice formation ($T_{ice} = -19.4°C$) no changes in interchain space were detected. The membrane thickness $d_l$ is not affected by ice and a decrease in $d$ occurs due to a decrease in $d_w$, so-called ice-induced dehydration of intermembrane space.

The dependence of $d$ on the temperature for a complete cooling process is presented in Fig. 3 for a binary DPPC/water system (circles), ternary DPPC/DMSO/water system with a mole DMSO fraction $X_{DMSO} = 0.05$ (squares), and for the same ternary system without solvent excess, 20% w/w solvent concentration (triangles). An abrupt decrease of $d$ by $\Delta d = 5.9 \pm 0.6$Å at $T_{ice} = -19.4°C$ for a binary DPPC/water system is the result of the ice-induced dehydration of intermembrane space at the moment of ice formation in a bulk solvent. The slope of curves correspond to the first stages of membrane crystallization. The membrane crystallization is accompanied with the dehydration of intermembrane space, for this type of dehydration the characteristic time is 12 hours which corresponds to slow water diffusion from intermembrane space to bulk solvent [11]. In the case of ice-induced dehydration, water diffusion from intermembrane space to bulk solvent actually occurs in no time, the characteristic time is ≤ 3min.

The temperature of ice formation in the system $T_{ice}$ is determined from WAXS diffraction with an accuracy of ± 1.5°C. The quantity $\Delta d$ is the difference in the membrane repeat distance at the temperature before ice formation, $T_{ice} − 3°C$, and the temperature after ice formation, $T_{ice} + 3°C$. The values of $\Delta d$ and the ice formation temperature $T_{ice}$ decrease as $X_{DMSO}$ increases (see Table 1). $\Delta d = 0.5 \pm 0.6$Å at $T_{ice} = -23.2°C$ for a system with $X_{DMSO} = 0.05$ (see Fig. 3). Above $X_{MSO} = 0.05$ changes in the value of $d$ were not established. For a dehydrated DPPC/DMSO/water system with $X_{DMSO} = 0.01$ and $X_{DMSO} = 0.05$, neither ice reflections nor an abrupt decrease in the value of $d$ were detected. In Figure 3 the curve for a dehydrated system with $X_{DMSO} = 0.05$ demonstrates only a small derivation of 0.014Å/°C connected with membrane crystallization at cooling.

Similar ice-induced dehydration was measured in binary DMPC/water and DSPC/water systems. Ice formation occurs in a DMPC/water system at $T_{ice} = -17.50°C$ and leads to an abrupt decrease of $d$ from $59.8. \pm 0.3$Å before ice formation (T= –14.5°C) to $d = 53.5 \pm 0.3$Å after the ice formation (T= –21.5°C). The diffusion of free water from intermembrane space to bulk water decreases the repeat distance by $\Delta d = 6.3 \pm 0.6$Å. For a binary DSPC/water system, the corresponding values are: $d = 67.3 \pm 0.4$Å at $T = -15.6°C$ before ice formation, $d = 62.2 \pm 0.3$Å at $T = -21.6°C$ after ice formation. Ice formation at $T_{ice} = -18.6°C$ creates ice-induced dehydration of intermembrane space with $\Delta d = 5.1 \pm 0.7$Å. Changes in the repeat distance $\Delta d$ at the moment of ice formation have about equal values for three types of lipid molecules (DMPC, DPPC, DSPC) within experimental errors.

The results for the ice-induced dehydration of DPPC/DMSO/water, DPPC/glycerol/water, and DMPC/DMSO/water systems are presented in Tables 1, 2, and 3. Each table presents the dependence of the ice formation temperature $T_{ice}$ and $\Delta d$ on the cryoprotector mole fraction. The value of $\Delta d$ becomes smaller as cryoprotector concentration increases. For the cryoprotector mole fractions above 0.05, the value of $d$ is not influenced by ice formation in bulk solvents.

It is a typical property of biological objects (not only membranes) to expel some part of water to a bulk solvent during freezing as a result of the coexistence of free and bound water in biological systems [12]. The presence of a cryoprotector decreases ice-induced dehydration. DPPC/DMSO/water and DPPC/glycerol/water systems demonstrate the similar properties in the conditions of cooling and ice formation. It is important to note that the influences of DMSO and glycerol on the phospholipid membrane structure are absolutely different [6, 7, 8, 9]. The different interaction of DMSO and glycerol molecules with the membrane surface and a similar property of ternary DPPC/DMSO/water and DPPC/glycerol/water systems at cooling give us a possibility to suppose that the properties of ternary phospholipid/cryoprotector/water systems mainly depend on the phase diagram of the cryoprotector/water systems [13] and the quantities of free and bound water in intermembrane space [7, 14, 15].

**4. CONCLUSION**

The temperature of ice formation in a ternary phospholipid/cryoprotector/water system in the case of phospholipids: DMPC, DPPC, DSPC and cryoprotectors: DMSO, glycerol mainly depends on the phase diagram of the cryoprotector/water system. A decrease of the membrane repeat distance by $\Delta d$ depends on the cryoprotector mole fraction. For the cryoprotector mole fraction 0.04 – 0.05, expulsion of water from intermembrane space to bulk solvent is negligibly small at the moment of ice formation. For the cryoprotector mole fraction higher than 0.05, the formation of ice does not influence the membrane structure.

The presence of ice is not detected in ternary phospholipid/cryoprotector/water systems without solvent excess. In the temperature interval used (down to –55.4°C) the bound water has not freezed.

**ACKNOWLEDGEMENTS**
The authors are grateful to Dr. Sylviane Lesieur (URA 1218 of CNRS) for her help in the sample preparation and T.B. Kiseleva (JINR, Dubna, Russia) for her assistance in the data treatment. One of us, M.A. Kiselev, acknowledges the allocation of CEA Saclay grant which made reported measurements possible.

Table 1.
The ternary DPPC/DMSO/water system with excess solvent. The dependence of the ice formation temperature $T_{ice}$ and $\Delta d$ on the DMSO mole fraction.

| DMSO mole fraction | 0.00 | 0.01 | 0.02 | 0.05 |
|---|---|---|---|---|
| $T_{ice}$, (°C) | −19.4 | −19.7 | −22.1 | −23.2 |
| $\Delta d$, (Å) | 5.9 ± 0.6 | 3.6 ± 0.6 | 1.9 ± 0.6 | 0.5 ± 0.6 |

Table 2.
The ternary DPPC/glycerol/water system with excess solvent. The dependence of the ice formation temperature $T_{ice}$ and $\Delta d$ on the glycerol mole fraction.

| Glycerol mole fraction | 0.00 | 0.01 | 0.03 | 0.05 |
|---|---|---|---|---|
| $T_{ice}$, (°C) | −19.4 | −20.6 | −24.9 | −27.2 |
| $\Delta d$, (Å) | 5.9 ± 0.6 | 4.3 ± 0.6 | 0.1 ± 0.6 | 0.0 |

Table 3.
The ternary DMPC/DMSO/water system with excess solvent. The dependence of the ice formation temperature $T_{ice}$ and $\Delta d$ on the DMSO mole fraction $X_{DMSO}$.

| $X_{DMSO}$ | 0.0 | 0.01 | 0.02 | 0.04 | 0.05 | 0.06 | 0.08 | 0.10 |
|---|---|---|---|---|---|---|---|---|
| $T_{ice}$, (°C) | −17.5 | −17.7 | −22.0 | −24.9 | −30.2 | −34.2 | −38.5 | −40.2 |
| $\Delta d$, (Å) | 6.3 | 2.3 | 2.6 | 0.3 | 0.3 | 0.0 | 0.0 | 0.0 |

**Figure captions:**

Fig. 1. The arrangement of water molecules in a phospholipid/water system. Water molecules are divided in three fractions: one is the free water molecules located in the center of intermembrane space, another is the bound water molecules located in intermembrane space near the polar heads of DPPC molecules, the third is the bound water molecules located in the spatial region of DPPC head groups.

Fig. 2a. The SAXS time resolved experiment, detector 1. The sequence of diffraction patterns from a binary DPPC/water system recorded with the acquisition time 1 min on cooling the sample from $T = 14.1°C$ to $T = -29.5°C$. The oscillations of the intensity in the center of the figure correspond to the direct beam absorption in the beam stop. Symmetrical to the center of the beam left and right peaks are correspond to the first and second order diffraction from multilamellar structure of DPPC. The measured $q$ interval is from $-0.3 Å^{-1}$ to $0.3 Å^{-1}$.

Fig. 2b. The WAXS time resolved experiment, detector 2. The sequence of diffraction patterns from a binary DPPC/water system recorded with the acquisition time 1 min on cooling the sample from $T = 14.1°C$ to $T = -29.5°C$. Small broad peaks at all temperatures correspond to the diffraction from hydrocarborn chains of DPPC. Three peaks at $T_{ice} = -19.4°C$ correspond to powder diffraction from hexagonal ice. The value of $q$ decreases from the left-hand side of this figure side to its right-hand side. The measured $q$ interval is from $1.9 Å^{-1}$ to $1.3 Å^{-1}$.

Fig. 3. The dependence of the DPPC membrane repeat distance $d$ on cooling from $T = 14.1°C$ to $T = -55.4°C$ at a cooling rate of $1.5°C/min$. The binary DPPC/water system (circles). The DPPC/DMSO/water system with a mole DMSO fraction of 0.05 (squares), and the same ternary system without solvent excess (triangles). An abrupt decrease of $d$ equals $5.9 \pm 0.6 Å$ for a binary system at the temperature of ice formation $T_{ice} = -19.4°C$.

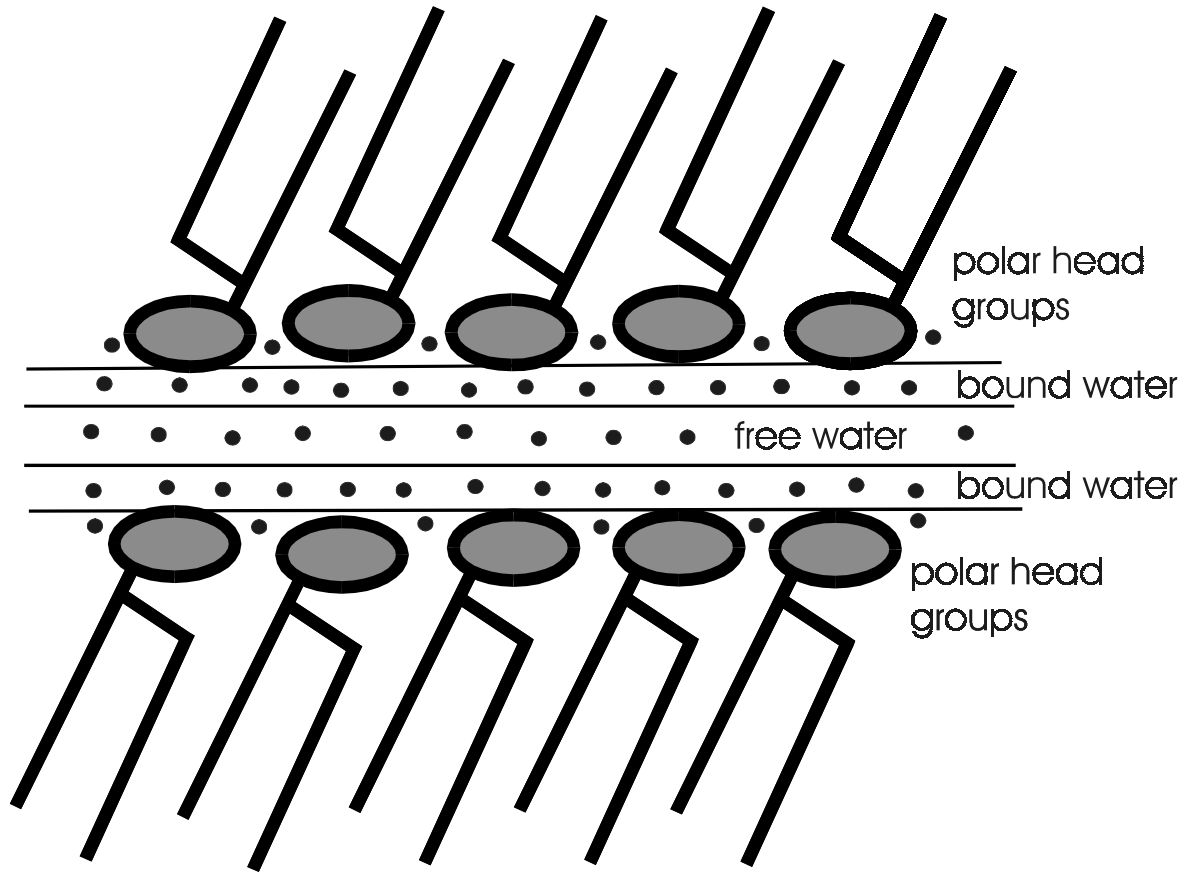

Figure 1.

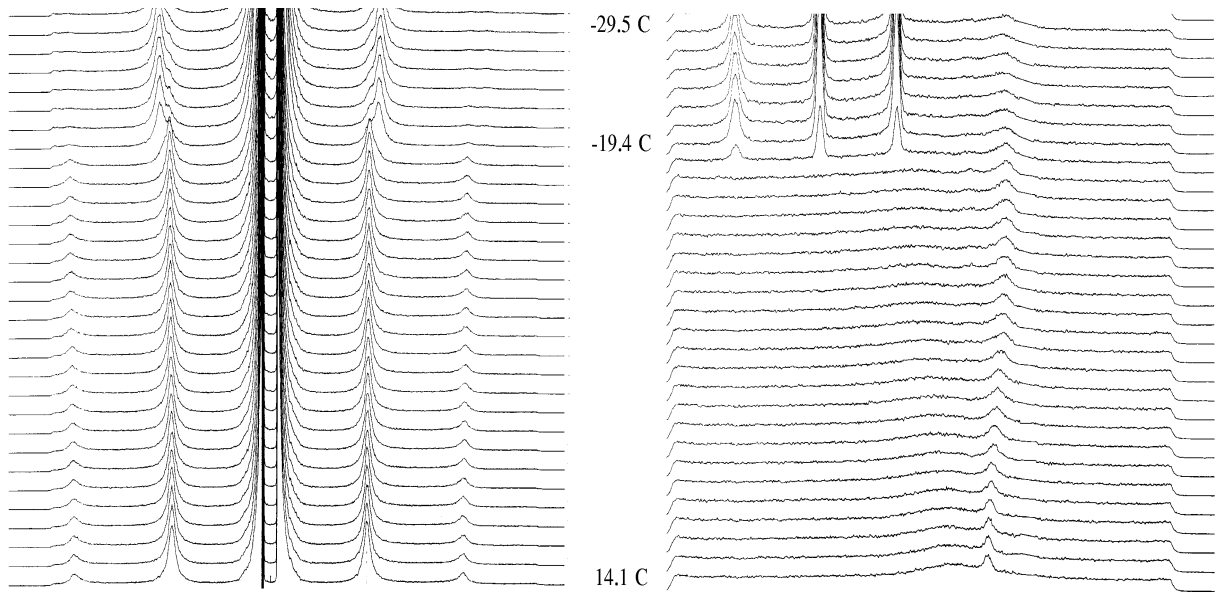

Figure 2a

Figure 2b

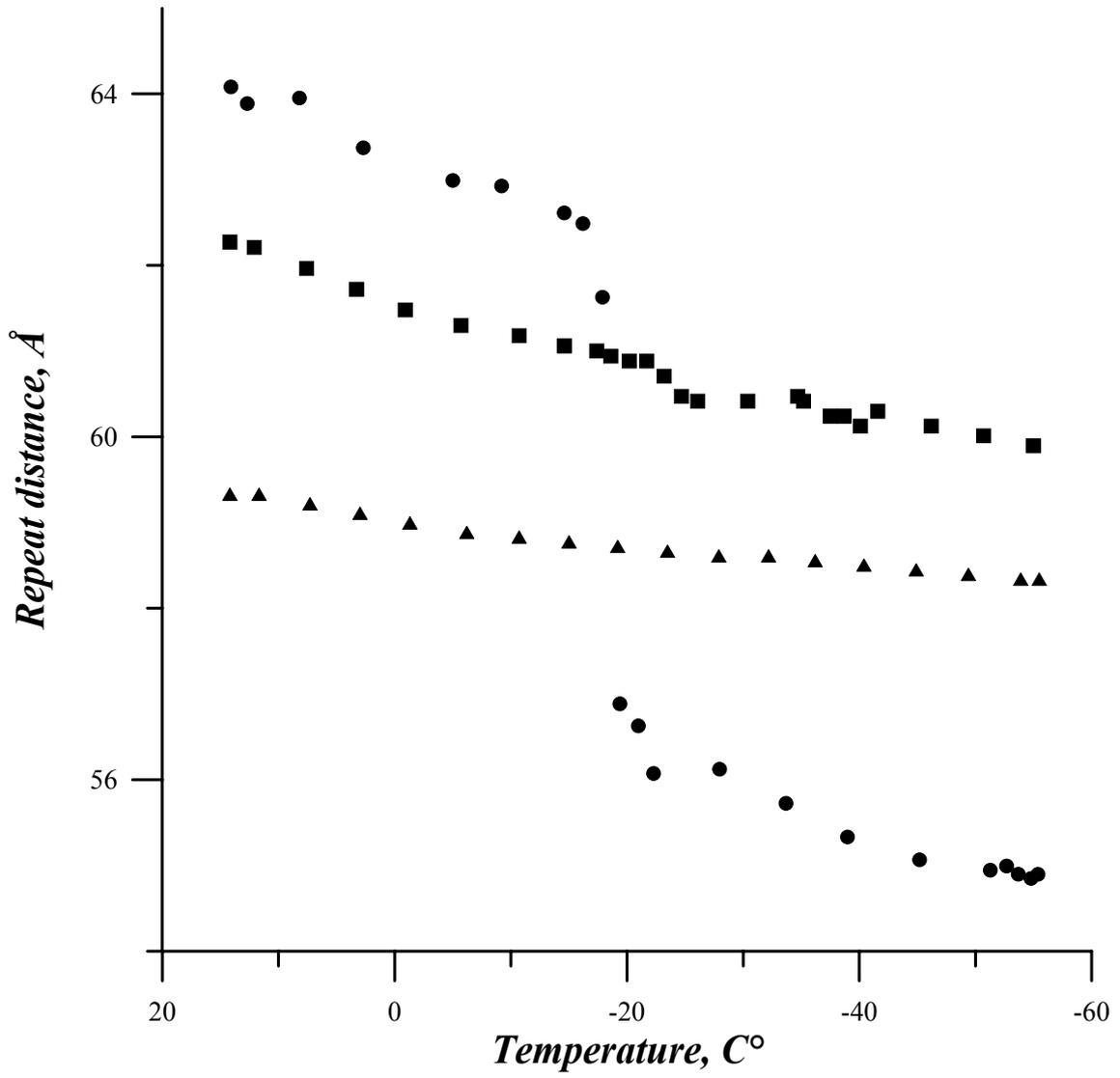

Figure 3